\documentclass[twocolumn]{openjournal}

\usepackage[english]{babel}
\usepackage[utf8x]{inputenc}
\usepackage[T1]{fontenc}
\usepackage{aas_macros}
\usepackage{graphicx}
\usepackage{amssymb}
\usepackage{setspace}

\shorttitle{Variability in SSTc2d J163134.1-240100}
\shortauthors{Scholz et al.}

\begin{document}

\title{Variability in SSTc2d J163134.1-240100, a brown dwarf with quasi-spherical mass loss}


\author{Aleks Scholz}
\affiliation{SUPA, School of Physics \& Astronomy, University of St Andrews, North Haugh, 
St Andrews, KY16 9SS, United Kingdom, email: as110@st-andrews.ac.uk}

\author{Koraljka Muzic}

\affiliation{Instituto de Astrofísica e Ciências do Espaço, Faculdade de Ciências, Universidade de Lisboa, Ed. C8, Campo Grande, 1749-016 Lisbon, Portugal}

\author{Victor Almendros-Abad}
\affiliation{Istituto Nazionale di Astrofisica (INAF) – Osservatorio Astronomico di Palermo, Piazza del Parlamento 1, 90134 Palermo, Italy}

\author{Antonella Natta}
\affiliation{School of Cosmic Physics, Dublin Institute for Advanced Studies, 31 Fitzwilliam Place, Dublin 2, Ireland}

\author{Dary Ruiz-Rodriguez}
\affiliation{National Radio Astronomy Observatory, 520 Edgemont Rd., Charlottesville, VA 22903, USA}

\author{Lucas A. Cieza}
\affiliation{Nucleo de Astronomia, Facultad de Ingenieria, Universidad Diego Portales, Av. Ejercito 441, Santiago, Chile}

\author{Cristina Rodriguez-Lopez}
\affiliation{Instituto de Astrofísica de Andalucía (IAA-CSIC), Glorieta de la Astronomía s/n, 18008 Granada, Spain}

\begin{abstract}
We report on a search for variability in the young brown dwarf SST1624 ($\sim$M7 spectral type, $M\sim 0.05\,M_{\odot}$), previously found to feature an expanding gaseous shell and to undergo quasi-spherical mass loss. We find no variability on timescales of 1-6\,hours. Specifically, on these timescales, we rule out the presence of a period with amplitude $>1\%$. A photometric period in that range would have been evidence for either pulsation powered by Deuterium burning or rotation near breakup. However, we see a 3\% decrease in the K-band magnitude between two consecutive observing nights (a 10\,$\sigma$ result). There is also clear evidence for variations in the WISE lightcurves at 3.6 and 4.5$\,\mu m$ on timescales of days, with a tentative period of about 6\,d (with a plausible range between 3 and 7\,d). The best explanation for the variations over days is rotational modulation due to spots. These results disfavour centrifugal winds driven by fast rotation as mechanism for the mass loss, which, in turn, makes the alternative scenario -- a thermal pulse due to Deuterium burning -- more plausible.
\end{abstract}


\section{Introduction} 
\label{sec:intro}

Photometric monitoring has proven to be a useful observational tool in the exploration of brown dwarfs. When they are young, substellar objects often show variability due to magnetic activity, accretion, and disks, analogous to the more massive T Tauri stars \citep{scholz2009,moore2019}. For evolved brown dwarfs, photometric variability can signify the presence and evolution of clouds in the atmosphere \citep[e.g.][]{metchev2015}. Variability studies have been able to measure substellar rotation periods as a function of age \citep{bouvier2014}, as well as constrain atmospheric properties and early evolutionary processes in brown dwarfs. 

Thanks to numerous monitoring studies in the past twenty years, we know that young brown dwarfs show rotation periods ranging from a few hours to several days, whereas old evolved brown dwarfs generally have periods shorter than one day \citep{moore2019,vos2022}. At young ages, the shortest measured periods are around the physical limit -- the period corresponding to breakup velocity, where centrifugal and gravitational forces are in balance at the equator. Rotation near or around breakup in young brown dwarfs has been reported by various authors, including \citet{zapatero2003,caballero2004,scholz2005,rodriguez2009}. Such fast rotation will have significant effects on internal structure and early evolution, which needs to be accounted for in models \citep{yoshida2023}.

\begin{figure}
\includegraphics[scale=0.38]{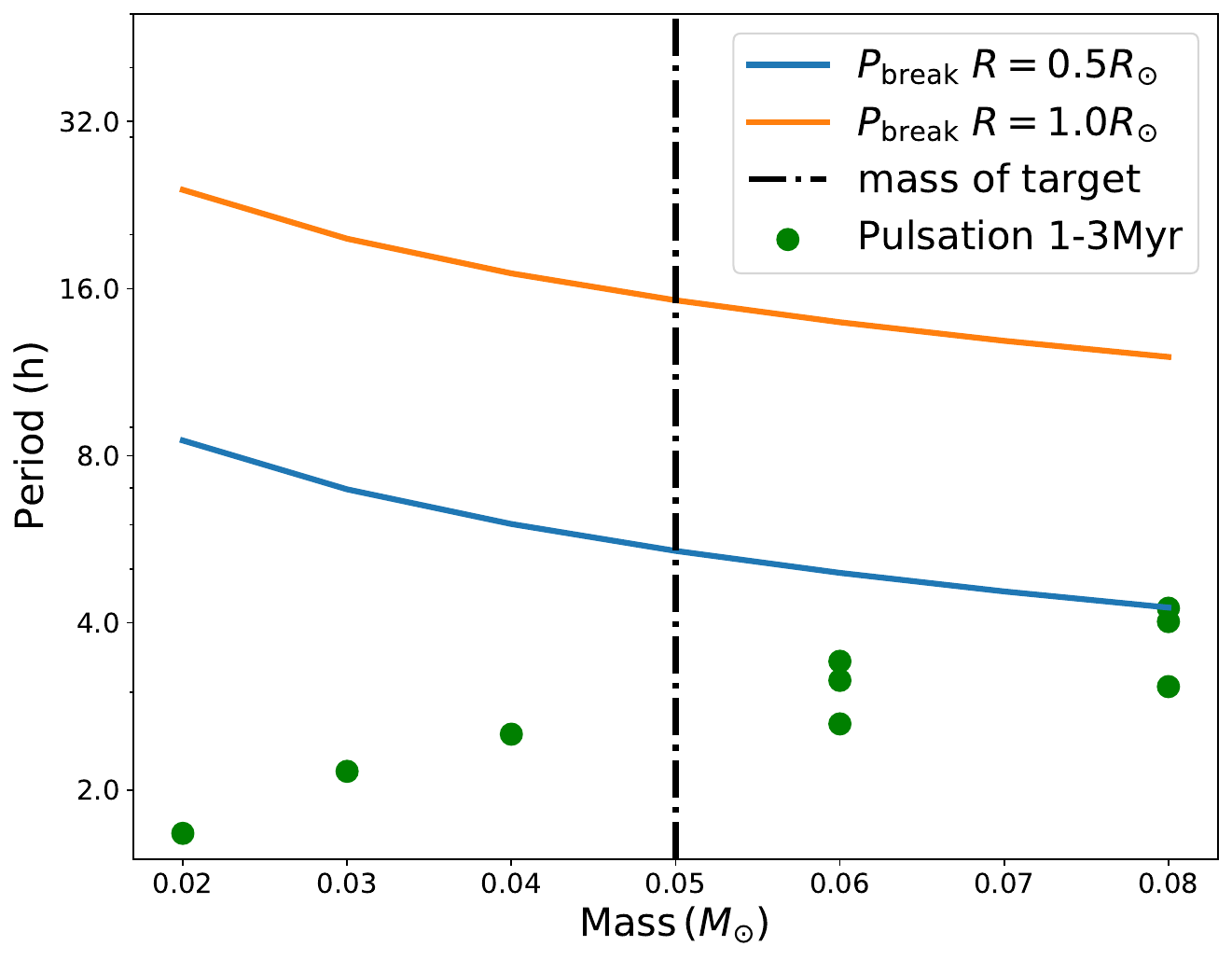}
\caption{Estimated rotation period at breakup velocity (lines) and pulsation period (dots) for young brown dwarfs, as a function of mass. The breakup periods are shown for two different radii, consistent with the expected values for inflated, young substellar objects. Pulsation periods are the values given by \citet{palla2005} for ages between 1 and 2\,Myr (their Table 1). For masses of 0.06 and 0.08$\,M_{\odot}$ these authors list several pulsation periods, for ages ranging from below 1 to 3\,Myr. The estimated mass of our target is marked by a dash-dotted line. \label{fig:rotpuls}}
\vspace{0.3cm}
\end{figure}

The breakup velocity is calculated using $v=\sqrt{2GM/3R}$, following \citet{porter1996}. The factor $2/3$ accounts for the fact that an object rotating at breakup will be oblate. For convenience, this corresponds to $P=0.14205 \times R^{1.5} / M^{0.5}$. For substellar masses and ages 1-5\,Myr, this yields breakup periods $>5$\,h, as shown in Figure \ref{fig:rotpuls} for two radii, typical for young brown dwarfs. As the objects age and contract, the breakup period drops with $\sim R^{1.5}$, but the rotation period (assuming angular momentum conservation) changes with $\sim R$, i.e. with increasing age the breakup period will drop further below the rotation period. As a result, field brown dwarfs which sometimes feature very short rotation periods of $\sim 1$\,h \citep{tannock2021} are still significantly above breakup.

Another physical process that can be probed by photometric monitoring is pulsations. Brown dwarfs exhibit Deuterium burning during the first few Myrs of their existence, predicted to lead to a luminosity burst \citep{salpeter1992} and to radial or non-radial pulsations.  \citet{palla2005} show in a stability analysis of Deuterium burning brown dwarfs the presence of fundamental modes with periods between 1 and 5\,h. For 0.1$\,M_{\odot}$ stars, \citet{lopez2012} find periods of modes excited by Deuterium burning of 4.2-5.2\,h. A dedicated observational search for pulsation periods down to mmag precision, by \citet{cody2014}, however, remained unsuccessful. In principle, some of the shortest photometric periods measured so far for brown dwarfs (see above) could be caused by pulsations. Most authors prefer, however, to interpret the periodic variability in young brown dwarfs as a result of rotation combined with magnetic surface spots. If pulsations were confirmed, it would open a new route to determine and ascertain fundamental properties in brown dwarfs during the phase of contraction.

In Figure \ref{fig:rotpuls} we include the estimated pulsation periods from \citet{palla2005} for ages of 1-2\,Myr and substellar masses. As can be appreciated from this figure, the gap between pulsation periods and breakup periods widens with decreasing mass and increasing radius. A period below 4\,h for such a young very low mass object would be very difficult to interpret as rotation period, offering a window to search for pulsations.

\section{The target}
\label{sec:target}

\begin{figure}
\includegraphics[scale=0.48]{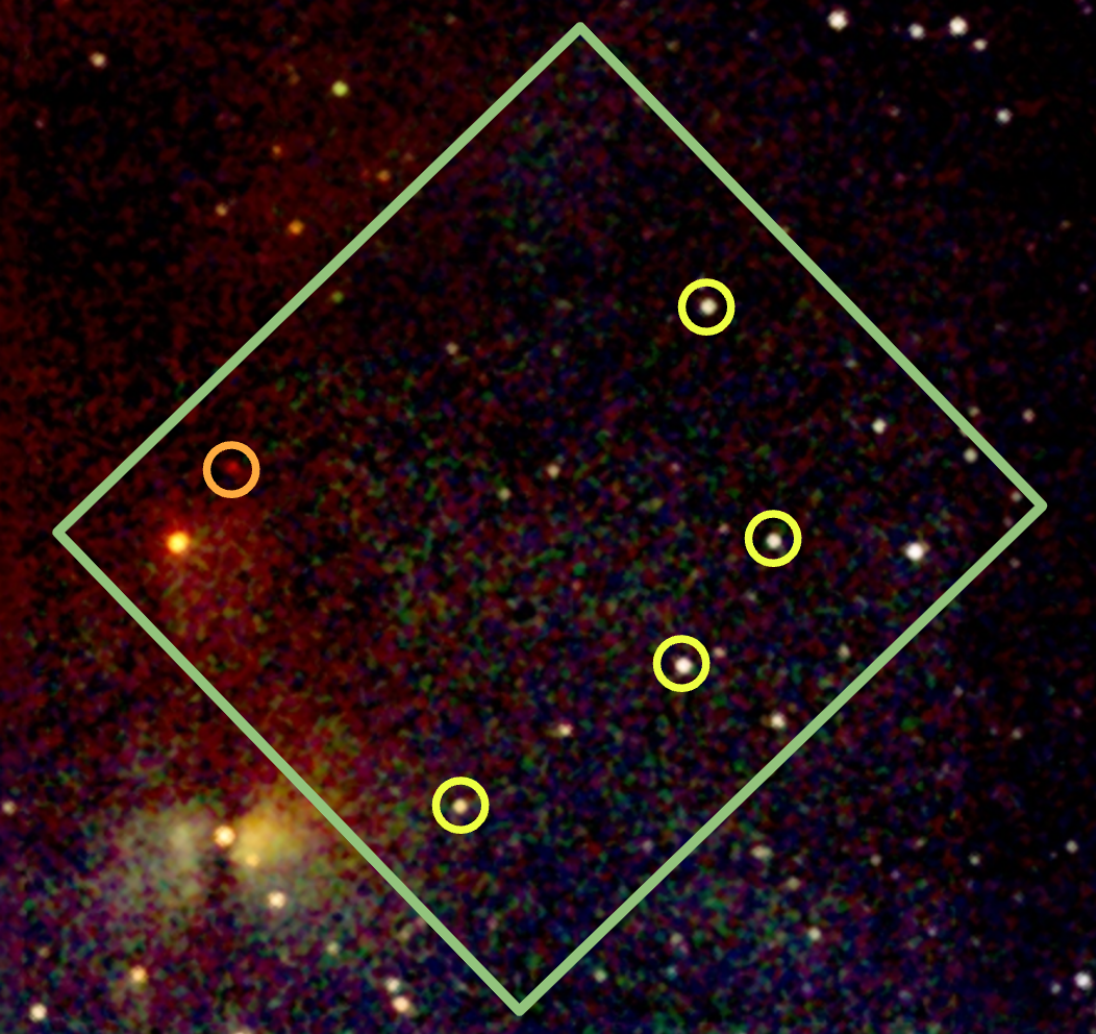}
\caption{2MASS JHK colour image of the field around our target. Marked in green is the $4.9'\times4.9'$ SOFI field. The target is marked in orange; the four stars used as comparison stars in yellow. North is up and east is left. \label{fig:field}}
\end{figure}

SSTc2d J163134.1-240100 (hereafter SST1624) is a young brown dwarf in the Ophiuchus star forming complex, identified originally in the Spitzer 'Cores to Disks' survey \citep{evans2009} as a faint, red young stellar object. A 2MASS image of the area around the target is shown in Figure \ref{fig:field}. In recent ALMA observations as part of the ODISEA program \citep{cieza2019}, it was found to have a spherical, oblate envelope with a diameter of 200-300\,AU, which shows the kinematics of an expanding shell \citep{ruiz2022}. The shell has been detected in $^{12}$CO, but not in the mm continuum. After correcting for extinction, the object has an infrared spectral energy distribution consistent with a diskless photosphere. This is very unusual -- so far very few young brown dwarfs have been detected in CO with ALMA, and the ones that are, usually have an obvious disk \citep{ricci2014}. On the other hand, numerous young brown dwarfs have been detected in the submm/mm continuum \citep{testi2016,sanchis2020}. We are thus confronted with an entirely new phenomenon that requires new ideas, continued exploration, and follow-up observations. 

\citet{ruiz2022} determine the nature of the source based on KMOS spectra, and are careful to rule out alternative explanations, such as an AGB star in the background or a collapsing molecular core. Proper motions confirm that the object is indeed a member of the young population in Ophiuchus, with an age of $\sim 2$\,Myr. Most likely, it is a 0.05$\,M_{\odot}$ brown dwarf, with mid/late M spectral type, overluminous, and seen through significant extinction of $A_V\sim42$\,mag. The best explanation for the gaseous expanding shell is strong, uncollimated mass loss. The age of the shell has been estimated kinematically to be $\sim 10^4$ years, a small fraction of the brown dwarf's age, and its gas mass (under plausible assumptions) is $1.5\,M_{\oplus}$. The $4\,\sigma$ upper limit in the continuum observations corresponds to a dust mass of $<1\,M_{\oplus}$.

\begin{figure*}
\includegraphics[scale=0.6]{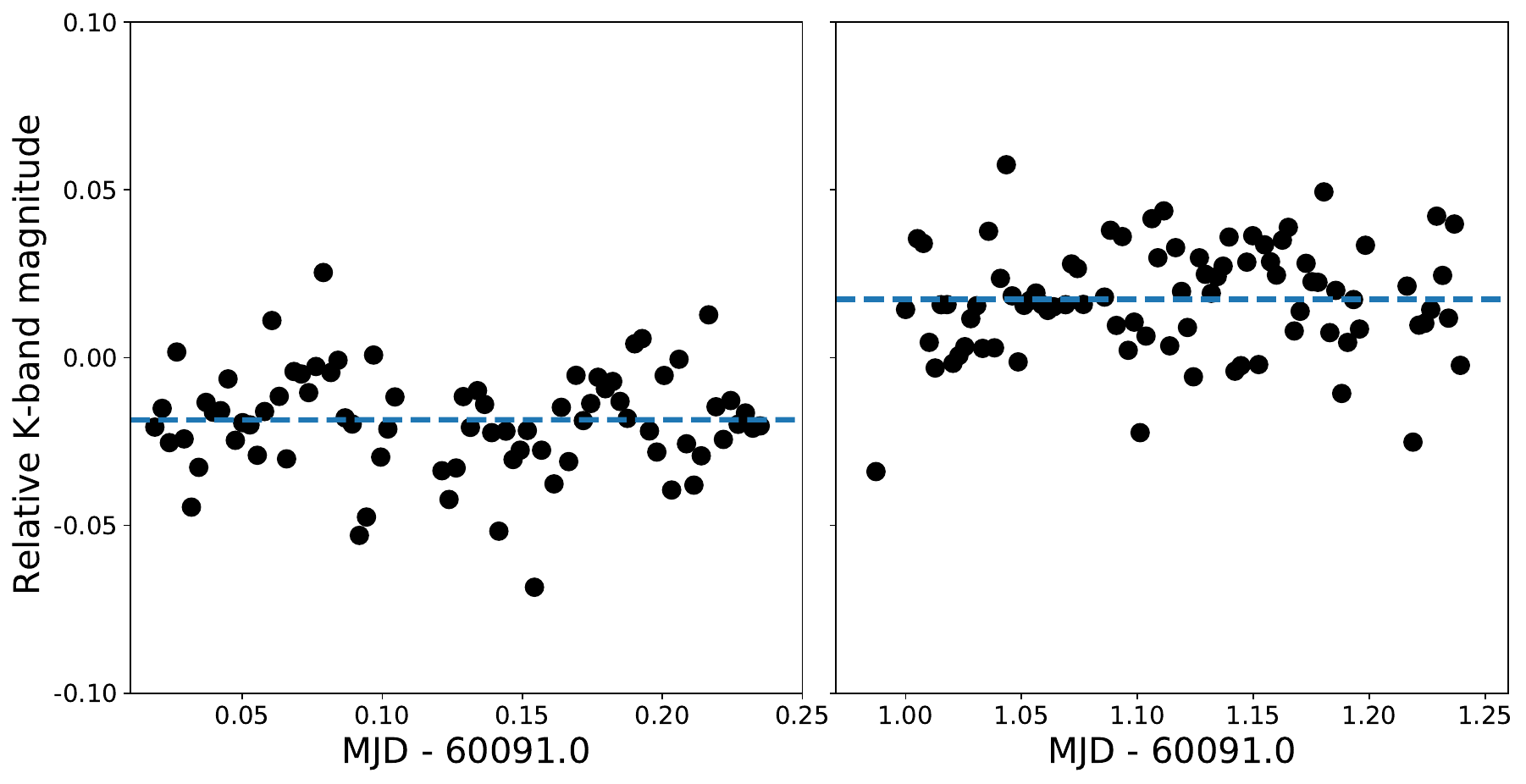}
\caption{K-band lightcurve for SSTc2d J163134.1-240100, measured over two nights in May 2023. The magnitudes are measured relative to the average of four field stars. The blue dashed lines mark the average for a given night. The typical error for individual datapoints is 0.01-0.02\,mag. \label{fig:kband}}
\vspace{0.9cm}
\end{figure*}

At this point the origin of the expanding shell is unknown. \citet{ruiz2022} suggest that the mass loss may be the result of a thermal pulse produced by the onset of Deuterium burning, analogous to the thermal pulses that drive mass loss in AGB stars \citet{kerschbaum2017}. Another option for a brown dwarf to eject a gaseous shell, originally suggested by \citet{scholz2005}, is through non-steady centrifugal winds caused by extremely fast rotation close to breakup speed -- analogous to fast rotating OB stars \citep{porter1996}. This would lead initially to a ring (a 'decretion disk'), expanding and dispersing to a quasi-spherical structure, consistent with the oblate shape of the shell of SST1624. 

As pointed out in Section \ref{sec:intro}, both pulsation and fast rotation can cause a periodic flux change detectable by photometric monitoring, with the plausible periods for pulsations being shorter than those for fast rotation near breakup (see Figure \ref{fig:rotpuls}). Testing for variability in SST1624 is therefore a promising way to explore the origin of its shell. This is the purpose of this paper.

For the record, \citet{ruiz2022} also discuss another option to explain the shell, the engulfment and consumption of a planetary companion. They deem this scenario as more exotic and less probable than the thermal pulse. As this idea would not produce a variability signature, we will not discuss it further here.

\begin{figure*}
\includegraphics[scale=0.6]{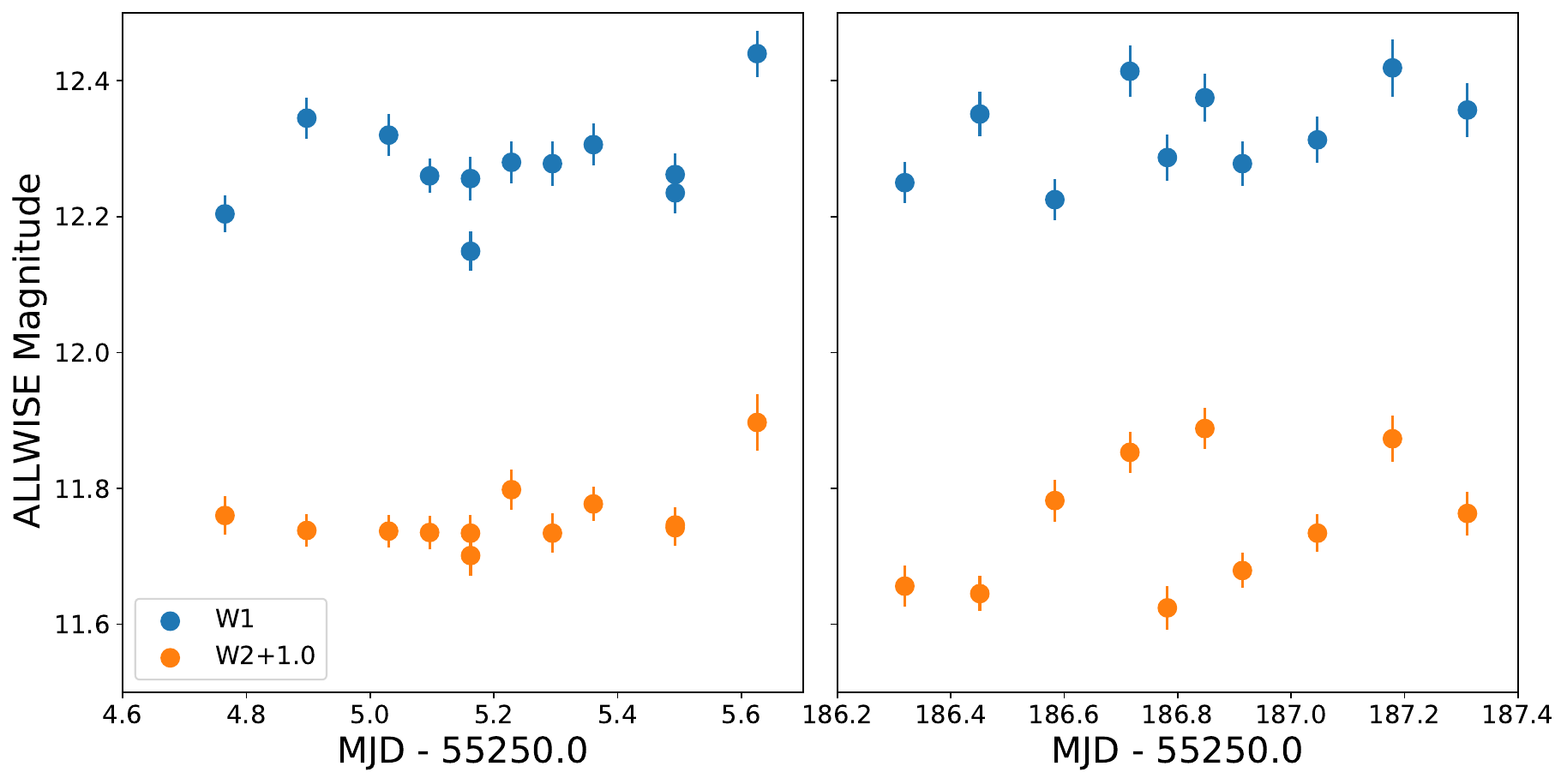}
\caption{ALLWISE lightcurve for SSTc2d J163134.1-240100, comprising 22 of the 23 datapoints in the archive. The W2 magnitudes are offset by 1.0\,mag for clarity. \label{fig:allwise}}
\vspace{0.9cm}
\end{figure*}

\section{Observations and data reduction}
\label{sec:obs}

\subsection{K-band monitoring}

We have monitored SST1624 over two nights, May 26th and May 27th 2023, with ESO/NTT and the infrared camera SOFI, under proposal ID 111.24GN. Overall, this observing run comprised three nights, but the third one was not usable due to clouds. In each of the two nights, we stayed on the target for 5-6\,h, using the $K_S$ filter, DIT of 20\,s, NDIT of 5 or 10, and employing random dithers. In both nights, the conditions were stable, with clear skies, low humidity (8-26\%), moderate wind speed (5-11\,ms$^{-1}$, well below the pointing limit of 14\,ms$^{-1}$ for this telescope), and temperature between 11 and 16 degrees Celsius, all according to the ESO Meteo monitor. The seeing in our data is between 1 and 2\,arcsec, without any rapid changes.

The data reduction includes cross-talk correction, flat fielding, sky subtraction, and bad-pixel correction. Individual exposures were combined, where necessary, to achieve a uniform total exposure time per frame of 200\,s. On each image, we performed aperture photometry using {\it photutils} \citep{bradley2023} using a constant aperture of 10 pixels, which typically corresponds to 2-3 FWHMs of the seeing-limited PSF. This includes background subtraction, where the background was measured in an annulus around the source. We opted not to adapt the aperture to the changing FWHM, to maintain consistency across the lightcurve. The aperture radius was chosen to optimise the signal-to-noise in the lightcurve, after experimenting with a wider range of radii.

The raw lightcurve for SST1624 and all other stars in the field show substantial variations due to the change in airmass over the course of the night. We chose four stars in the same field that are 2-3\,mag brighter than our target (marked yellow in Figure \ref{fig:field}), averaged their lightcurves, and subtracted this average from the target lightcurve (in units of magnitudes) to correct for atmospheric effects. The final lightcurve for SST1624 for those two nights is shown in Figure \ref{fig:kband}. Judged by the noise in the comparison star's lightcurves, the typical error for individual photometry points for the target is 0.01-0.02\,mag (or 1-2\%). 

We made sure that the lightcurve does not depend significantly on the choice of comparison stars, by checking the result with only a subset of them. We also checked the comparison stars against each other, and do not find any signs of variability in them. We note that the same calibration process also corrects for other environmental influences on the photometry, like changes in seeing. Using the comparison stars for calibration, we determine that SST1624 has a K-band magnitude of $14.08\pm 0.05$ in the 2MASS system, consistent with its 2MASS value of 14.06.

\subsection{Archival WISE lightcurves}

We also obtain the WISE lightcurve at 3.6 (W1) and 4.5$\,\mu m$ (W2) for SST1624 from the 'ALLWISE Multiepoch Photometry Table' \citep{cutri2021} and the 'NEOWISE-R Single Exposure (L1b) Source Table' \citep{mainzer2014}. The ALLWISE table contains 23 epochs for this source within 200\,d, starting at MJD55254. With one exception, all of these measurements are concentrated in two 24\,h spans, separated by $\sim 180$\,d, and shown in Figure \ref{fig:allwise}. 

\begin{figure}
\includegraphics[scale=0.38]{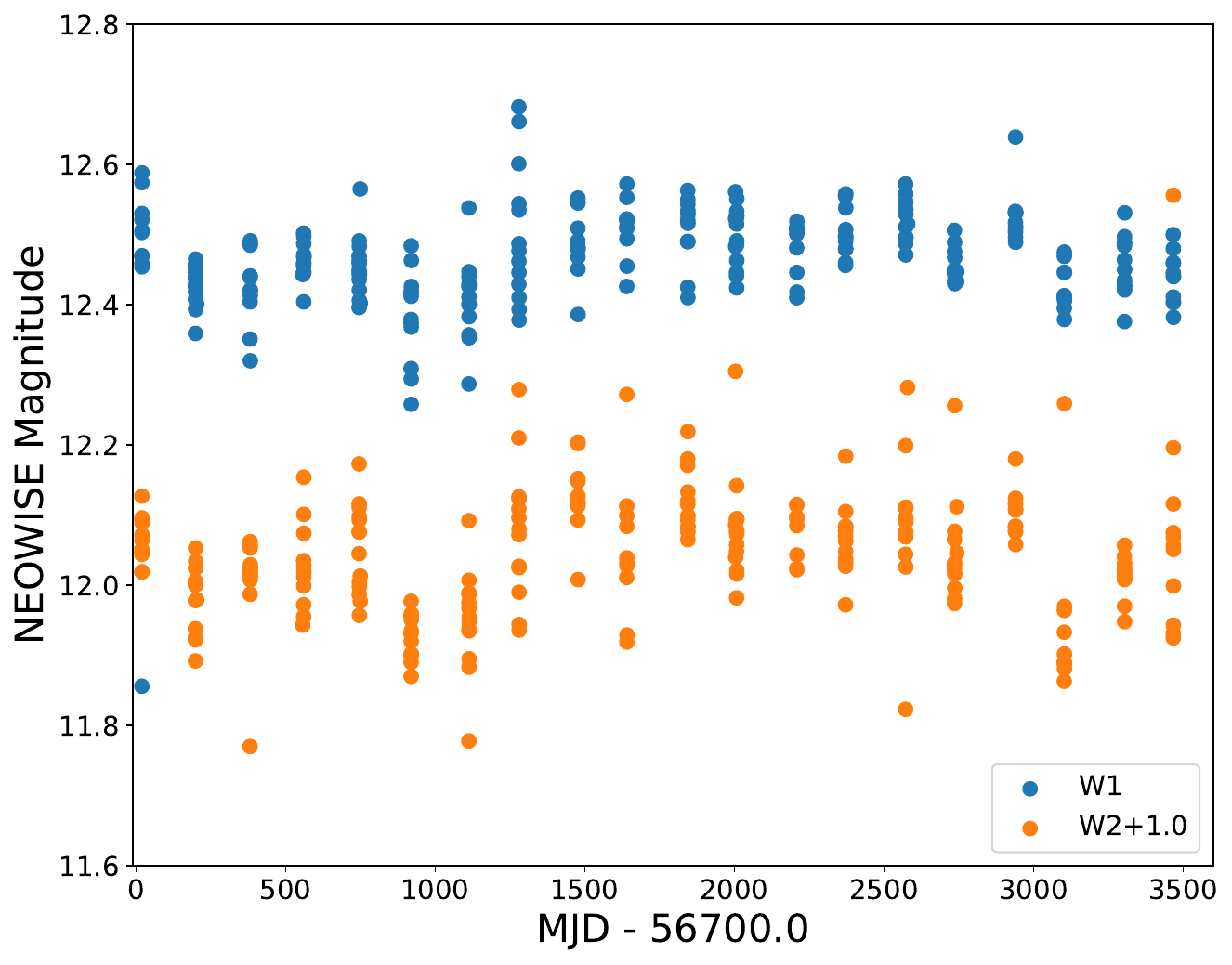}
\caption{NEOWISE-R lightcurve for SSTc2d J163134.1-240100, comprising 237 datapoints. A few outliers are not shown here. The W2 magnitudes are offset by 1.0\,mag for clarity. The typical photometric error is 0.03\,mag. \label{fig:neowise}}
\vspace{0.3cm}
\end{figure}

NEOWISE-R provides 237 good detections (as of April 2024), over a duration of $\sim 3500$\,d, starting at MJD 56700.0, with 20 day-long spans covered by about 10 datapoints each. Again the measurements are available for the two bands W1 and W2. We note that this lightcurve ends at MJD 60167 which corresponds to August 2023, a few months after our dedicated K-band monitoring. There are, however, no datapoints coinciding with our observations. The full lightcurve is shown in Figure \ref{fig:neowise}.

Both the ALLWISE and NEOWISE archives provide errorbars for individual photometry points. Those errors are determined from an assessment of the repeatability of the measurements as a function of magnitude and galactic latitude \citep{wright2010}. The nearest source is a bright star 36\,arcsec to the southeast. This source is the reason the WISE archive reports a risk of contamination. The photometric quality in the ALLWISE photometry is robust with a signal-to-noise ratio $>10$ (i.e. the {\it ph\_qual} flag is 'AAAA').

\section{Lightcurve analysis}

\subsection{The Max-F test}
\label{sec:maxf}

For the analysis presented in this paper, we devised a simple, robust test to check for the presence of a period in noisy data by comparing the variance in the lightcurve before and after subtracting a periodicity, a procedure hereafter called {\it Max-F}. In short, the routine tests to what extent the noise in the lightcurve could be caused by a periodicity with specific properties.

We define a range of test periods for a given dataset. For each test period we calculate the phases of the datapoints (between 0.0 and 1.0). Then we subtract sinecurves of the given test period, varying the amplitude and zeropoint of the phase. The latter is changed from 0.0 to 1.0 with a stepsize of 0.01. We record the standard deviation of the lightcurve for each subtracted sinecurve, $\sigma_{P,i}$, and the variance ratio $F= \sigma^2 / \sigma_{P,i}^2$. Here $\sigma$ is the standard deviation of the original lightcurve. We find the maximum value for $F$ for each period, and plot that as a function of period. 

This procedure gives us a measure of the maximum fraction of the noise that could be explained by the presence of a sinusoidal period. The routine constitutes in essence an {\it F-test for the equality of variances}, a standard statistical test. A similar test -- but only to verify a previously adopted period -- was used for example by \citet{scholz2004} in the lightcurve analysis of young very low mass sources. For normally distributed datasets, the quantity $F$ should follow an F-distribution (hence the name). 

\begin{figure*}
\includegraphics[scale=0.6]{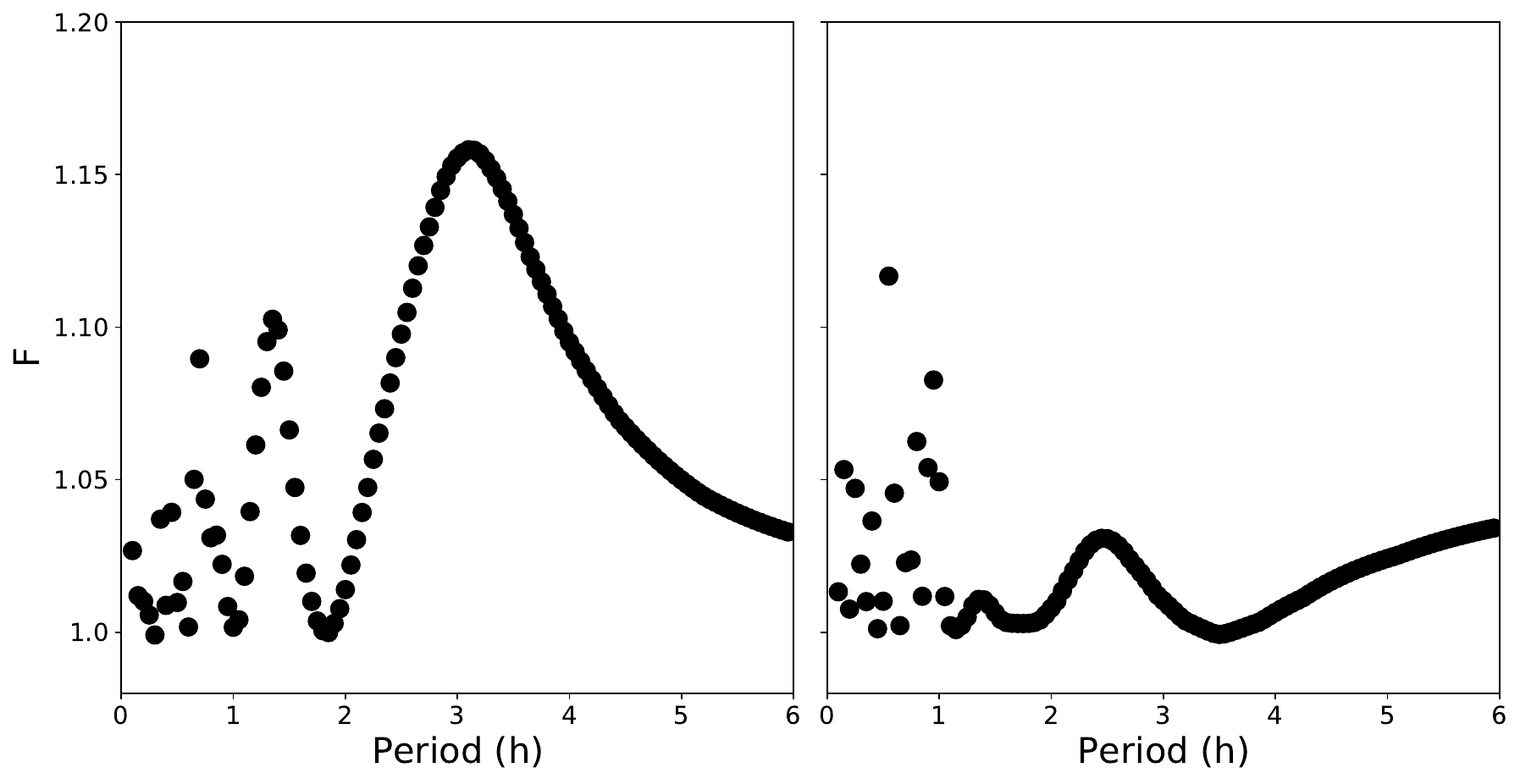}
\caption{Results of the {\it Max-F} test for our K-band lightcurve, for night 1 on the left, and night 2 on the right. The figure show the value of $F$ as a function of test period, as described in Section \ref{sec:maxf} and \ref{sec:kbandana}. \label{fig:maxf_k}}
\vspace{0.9cm}
\end{figure*}

\subsection{K-band lightcurve}
\label{sec:kbandana}

The lightcurve we obtained from the SOFI observations appears unremarkable. By eye, it shows noise around a flat mean, in each of the two consecutive nights, as can be appreciated from Figure \ref{fig:kband}. There is no obvious structure, in particular no obvious periodic variability. The standard deviation is 1.5 and 1.6\%, for night 1 and night 2 respectively. 

To check for gradual trends, we fit the data in individual nights with a straight line. In the first night, the best fitting slope is -0.006\,mag/day, which is consistent with zero. In the second night, the fit gives 0.026\,mag/day or 0.010\,mag/day when ignoring the very first outlying datapoint. Again this would be consistent with zero over a time window of 6\,h. The error in the slope for both nights amounts to 0.03\,mag/day, considerably larger than the slope itself, as expected for a null result.

There is, however, a significant offset between the mean magnitude in the two nights; the object dimmed by 3\% over the course of 1\,d, a 10\,$\sigma$ detection given the error of the mean in the lightcurves in individual nights. This shift is not seen in individual comparison stars; the difference between the nightly averages for the four comparison stars is 0.55\%, 0.12\%, 0.60\%, and 0.07\%.

Given that the observing setup was identical in both nights, and the range of airmasses was the same, too, this dimming is likely caused by intrinsic variability in our target. We note that this offset, in combination with the linear slopes measured for individual nights, could be consistent with a sinusoidal variation  -- in that case the object would be near maximum in the first night; the period would be on the order of days.

We carried out the {\it Max-F} test described in Section \ref{sec:maxf}, for a period range from  0.1\,h to 6\,h in steps of 0.05\,h. The amplitudes were varied between 0.001 and 0.03\,mag, in steps of 0.001\,mag. The results are shown in Figure \ref{fig:maxf_k}. For the first night, we find that the maximum value of $F$ is 1.16, for a period of 3.0\,h. For the second night, the maximum $F$ is 1.12, for periods shorter than 1\,h; for longer periods the maximum is 1.04.
For comparison, for $N\sim 80$, a F-value of $>1.45$ would mean that the two variances before and after subtracting the sinewave are significantly different with a false alarm probability $<5$\%. Taken together, these results are consistent with the variances being equal before and after period subtraction. It also means that at most a few percent of the variance can be explained by a potential period. 

To test whether or not an actual period with an amplitude $>0.01$\,mag (1\%) can be recovered with this method, we injected artificial periods into our lightcurve. We chose a range of periods from 1 to 5\,h, and a constant amplitude of 1\%. For those parameters, the maximum $F$ would be $\sim 1.3$, depending slightly on period, with a clearly defined maximum at the injected period value. The injected period is then also visible by eye in the lightcurve. This demonstrates that periods with these characteristics are easily recovered by our {\it Max-F} test. We can therefore rule out the presence of periods of 1-5\,h with amplitudes larger than 1\% in the K-band lightcurve.

\begin{figure*}[t]
\includegraphics[scale=0.6]{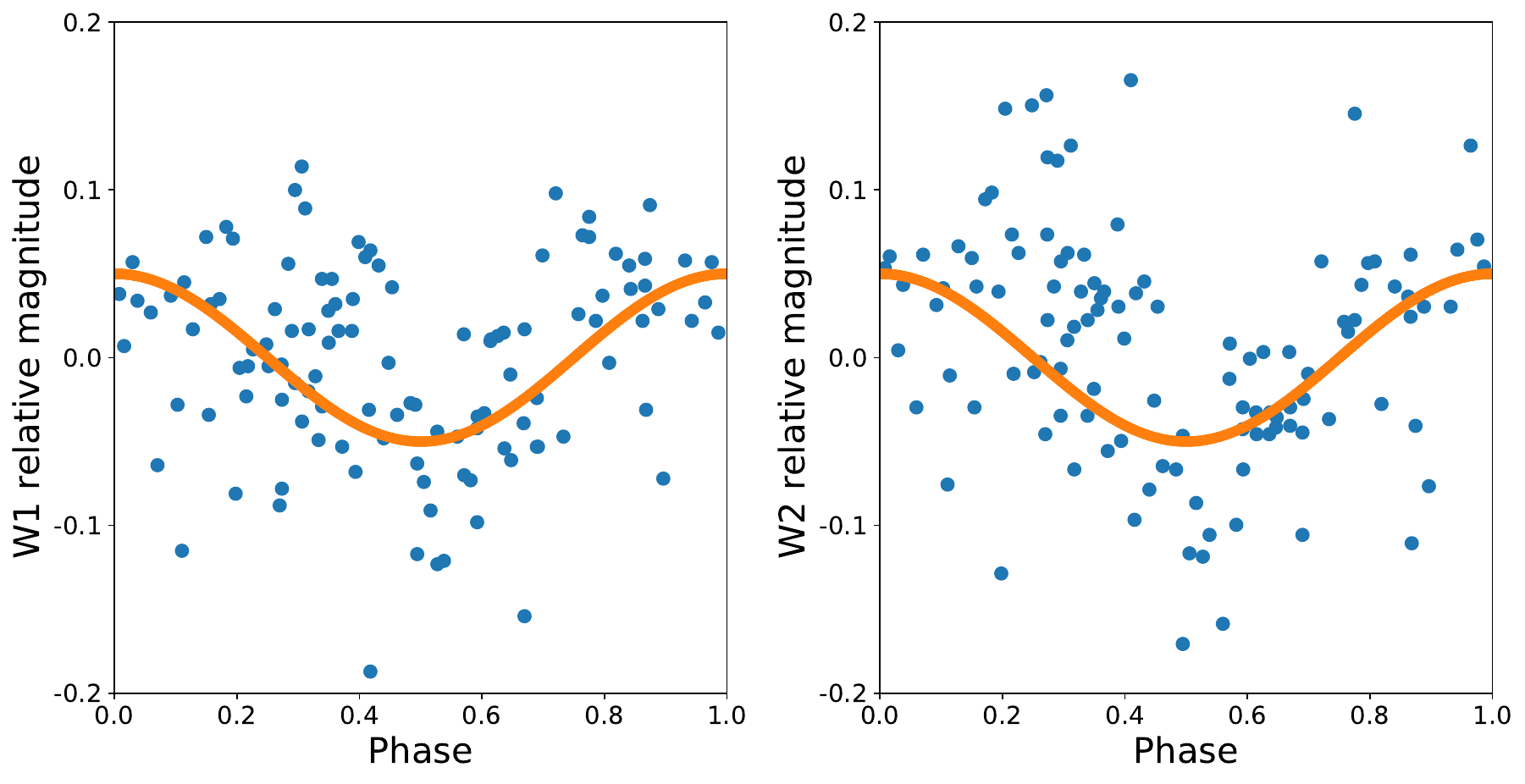}
\caption{The first half of the NEOWISE-R dataset, plotted in phase to a period of $P=6.0\,d$. Overplotted is the sinecurve with an amplitude of 0.05\,mag to guide the eye. The typical photometric error is 0.03\,mag. \label{fig:period}}
\vspace{0.9cm}
\end{figure*}

\begin{figure*}
\includegraphics[scale=0.6]{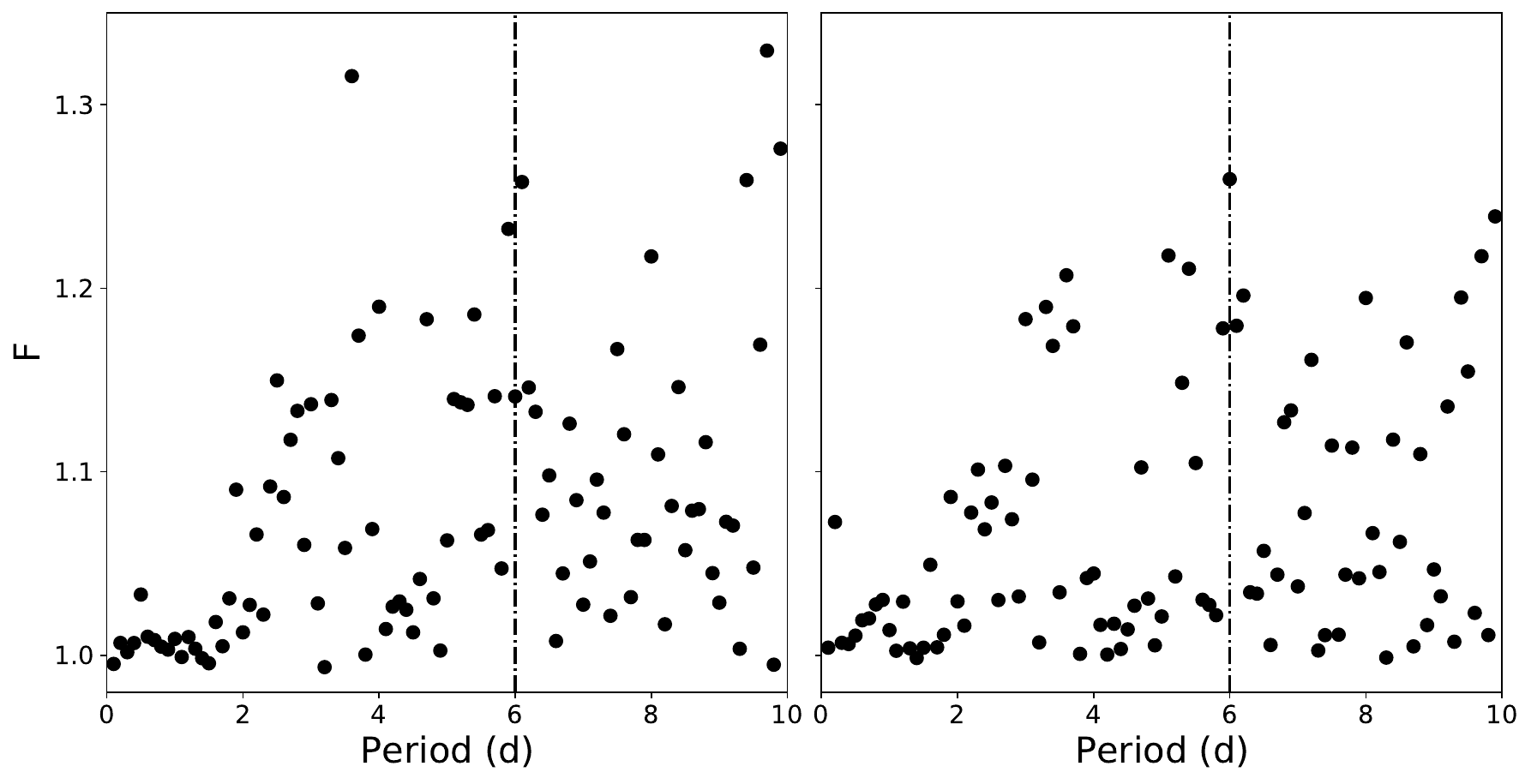}
\caption{Results of the {\it Max-F} test for the NEOWISE lightcurve (W1 on the left, W2 on the right; in both cases only the first half of the datapoints was used). The figure show the value of $F$ as a function of test period, as described in Section \ref{sec:maxf} and \ref{sec:wiseana}. \label{fig:maxf_wise}}
\vspace{0.9cm}
\end{figure*}

\subsection{WISE lightcurves}
\label{sec:wiseana}

In the ALLWISE and NEOWISE-R lightcurves, SST1624 shows clear variations, dominated by modulations on timescales of days, with typical peak-to-peak amplitudes of 0.1-0.3\,mag (see Figures \ref{fig:allwise} and \ref{fig:neowise}). These variations are significantly larger than the typical error, which is 0.03\,mag for a single epoch. Therefore it is plausible to assume the source is variable in the WISE lightcurves. Taken the two datasets together, the variability is sustained over more than 13 years. Over long timescales, the brightness of this object appears to be mostly stable within the margins of the intra-night variations.

The NEOWISE-R dataset allows a more detailed analysis. It consists of 20 sets of points, each measured within a couple of days, and separated by about 180\,d (see Figure \ref{fig:neowise}). Given this sampling, the NEOWISE-R lightcurve is not sensitive to the short timescales covered with our K-band monitoring, but can be useful to pick up variability on timescales ranging from days to years.

We carried out the same {\it Max-F} test we described in Section \ref{sec:maxf}, to search for a possible periodicity. For periods ranging from 1 to 10\,d, and for the full NEOWISE-R lightcurves, the maximum $F$ is 1.08 in W1 (for a period of 6\,d) and 1.12 in W2 (for a period of 8\,d). We noticed that the maximum $F$ increases significantly if we run the test only on the first half of the dataset. For band W1, we obtain peaks at periods of 3.6 and 6.1 and $>9.0$\,d, with $F>1.25$. For W2, the highest peak is for $P=6.0$\,d, with $F\sim 1.25$. The results for the first half of the dataset are shown in Figure \ref{fig:maxf_wise}. As a reminder, this means that the variance is reduced to 80\% ($1/1.25$) when subtracting this specific periodicity. However, for this value of $F$ there is still a substantial probability (11\%) that the two variances before and after subtracting the period are equal (assuming normally distributed samples).

In Figure \ref{fig:period} we show the first half of the NEOWISE-R datapoints plotted in phase for $P=6.0$\,d. As can be appreciated from this figure, the periodicity looks plausible by eye. It is therefore conceivable that parts or all of the variability seen in WISE data is caused by a flux modulation with a period of about 6\,d and an amplitude of $\pm 0.05$\,mag. We show the {\it Max-F} results in Figure \ref{fig:maxf_wise}, with the putative period of 6.0\,d marked.

\begin{figure*}[t]
\includegraphics[scale=0.6]{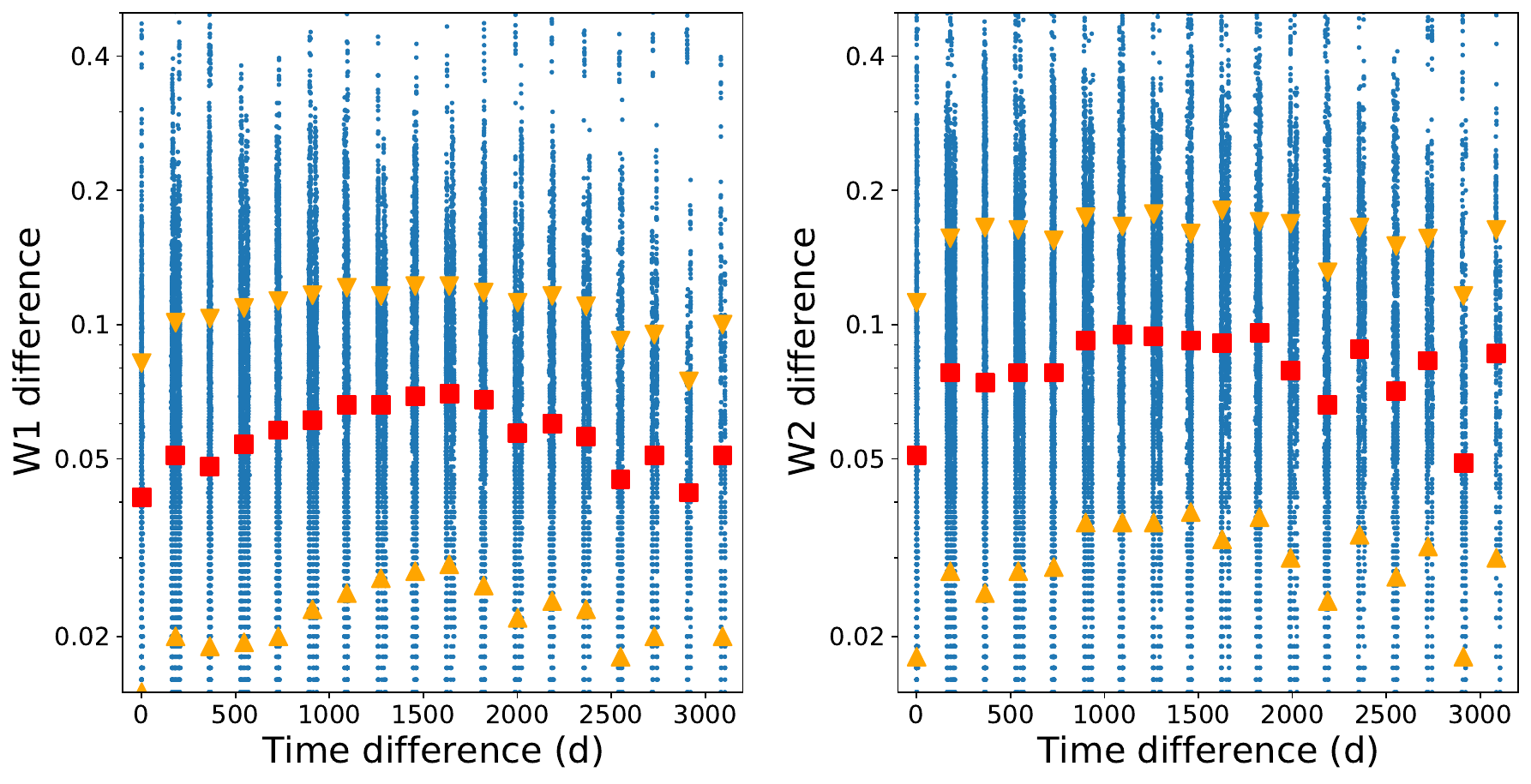}
\caption{Differences between NEOWISE-R magnitudes, as a function of time difference, for the two bands separately (W1 in the right panel, W2 in the left panel). The red squares show the median for an interval of $\pm 10$\,d around a given time difference. Similarly, the upwards and downwards pointing orange triangles show the 20\% and 80\% percentiles for the same intervals. The typical error in the magnitude difference is 0.04\,mag. \label{fig:wisediff}}
\vspace{0.9cm}
\end{figure*}

To quantify typical timescales of variability from the NEOWISE-R lightcurve, we calculated the difference between each pair of datapoints $\Delta M$, and plot this over their time difference $\Delta t$ (see Figure \ref{fig:wisediff}). This analysis was carried out for the two bands separately, but they both give very similar results. As can be seen in the figure, the overall range of $\Delta M$ is not a strong function of time difference $\Delta t$. The median of $\Delta M$ (shown as orange squares) is 0.04-0.05 for the shortest $\Delta t$ (hours to days), slightly larger than the photometric error of 0.03\,mag. The median of $\Delta M$ increases gradually to 0.07-0.10 up to $\Delta t \sim 1000$\,d, where it plateaus. The 80\% percentile of $\Delta M$ (shown with orange triangles pointing downwards) is by and large constant at 0.1\,mag for W1 and close to 0.2\,mag for W2. Overall, there is no strong evidence for long-term variability from the NEOWISE-R dataset.

\section{Summary and discussion}

SST1624 is a brown dwarf with quasi-spherical mass loss, a unique object that may give us insights into important aspects of the early evolution of substellar objects. Two possible mechanisms to explain the mass loss are a thermal pulse from the onset of Deuterium burning or centrifugal winds due to very fast rotation. In both cases, we may expect variability in short timescales, either due to pulsation or rotation. 

We present here infrared lightcurves for SST1624 from newly obtained high-cadence K-band ESO/NTT observations and from the publicly available ALLWISE and NEOWISE-R archives. Overall, there is clear evidence for variability in SST1624, on timescales of $>6$\,h to days, measured at 2-4.5$\,\mu m$ in all available datasets. On the other hand, we rule out variability, and specifically the presence of a photometric period on timescales $<6$\,h, for amplitudes $>1$\%. We do identify a tentative photometric period of 6\,d (with a plausible range of 3-7\,d) in the first half of the NEOWISE-R dataset, seen at 3.6 and 4.5$\,\mu m$. There may be additional variability on timescales of years. 

We note that our target was included in the variability search by \citet{park2021} using the first 6.5\,yr of NEOWISE data -- no variability was found. This study, however, was only focused on long-term variations over timescales longer than days. Also, the NEOWISE-R dataset is now significantly longer. 

The variability on day-long timescales, and the tentative period of 6\,d, is best explained by magnetic activity, i.e. spots on the surface co-rotating with the object. This implies that the rotation period for this object is in the order of a few days. Period and amplitude are comparable to what has been found for young brown dwarfs -- periods typically range from a few hours to several days, amplitudes are 2\% or larger \citep{scholz2005,moore2019}. The WISE lightcurve is very clumpy, with gaps of 6 months between clumps of data. Over such long timescales, the spot distribution can change, leading to phaseshifts and amplitude changes. For timescales around 1\,d, the number of datapoints is very low. These two characteristics in combination hamper any period search and can easily explain why the period is not apparent in the entire dataset. To robustly confirm the tentative period of 6\,d as the rotation period, a dedicated monitoring campaign over at least one week, if possible without daytime gaps, would be needed. 

The only other conceivable explanation for the observed variability is the presence of some residual circum-sub-stellar dust close to the object that causes variability through obscuration \citep{scholz2009} -- however, the lack of an infrared excess and the lack of a mm continuum detection renders this interpretation unlikely. 

Coming back to the topic of the quasi-spherical mass loss, the unique feature of this object: With our dedicated K-band monitoring, we set out to find periods on timescales of hours, to distinguish between pulsations (1-4\,h) and rotation near breakup ($>5$\,h). Since we did not find periods on these timescales, the available data does not give us a definitive answer on the cause of the mass loss. Having said that, the fact that there is clear variability over timescales of days, best explained by rotational modulations, would imply that the object is in fact not rotating close to breakup. Rotation periods of several days are at the high end of the period distribution for young brown dwarfs, see Figure 13 in \citet{moore2019}. This renders the scenario where the mass loss is due to the onset of Deuterium burning, favoured and worked out by \citet{ruiz2022}, more likely. The absence of a pulsation period in the lightcurves could mean that the object has stopped pulsating since the ejection of the shell, or that the pulsations happen at amplitudes below 1\%, as it is likely the case for pulsating M dwarfs \citep{rodriguez2016,rodriguez2019}. 

This brown dwarf deserves further time-series observations to a) verify the rotation period and b) search for pulsations with higher precision. As the first brown dwarf with observed quasi-spherical mass loss, this source may present an opportunity to learn more about the hitherto unexplored aspects in the early evolution of brown dwarfs. Deuterium burning is a common feature in all young brown dwarfs, and if this leads to mass loss, either sustained, or in bursts, it would affect the substellar mass function and angular momentum evolution. Exploring the prevalence and duration of substellar mass loss is an important task for future work.

\begin{acknowledgments}
We thank the referee for a prompt and thorough report on this paper.
We thank the ESO team at the NTT for the support prior and during the observing run. This publication makes use of data products from the Wide-field Infrared Survey Explorer, which is a joint project of the University of California, Los Angeles, and the Jet Propulsion Laboratory/California Institute of Technology, funded by the National Aeronautics and Space Administration. AS acknowledges support from the UKRI Science and Technology Facilities Council through grant ST/Y001419/1/. KM acknowledges support from the Fundação para a Ciência e a Tecnologia (FCT) through the CEEC-individual contract 2022.03809.CEECIND and research grants UIDB/04434/2020 and UIDP/04434/2020.
\end{acknowledgments}

\end{document}